\documentclass[10pt, conference, compsocconf]{IEEEtran}
\usepackage{graphicx}
\usepackage{amsmath}
\usepackage{subfig}
\usepackage{rotating}
\usepackage{multirow}
\usepackage{array}
\usepackage{algorithm}
\usepackage{algorithmic}

\begin{document}

\title{Monitoring and Controlling Power using\\ Zigbee Communications}

\author{N. Javaid, A. Sharif, A. Mahmood, S. Ahmed$^\sharp$, U. Qasim$^{\ddag}$, Z. A. Khan$^{\S}$\\

        COMSATS Institute of IT, Islamabad, Pakistan. \\
        $^\sharp$Mirpur University of Science and Technology, AJK, Pakistan.\\
        $^{\ddag}$University of Alberta, Alberta, Canada\\
        $^{\S}$Faculty of Engineering, Dalhousie University, Halifax, Canada.
        }

\maketitle

\begin{abstract}
Smart grid is a modified form of electrical grid where generation, transmission, distribution and customers are not only connected electrically but also through strong communication network with each other as well as with market, operation and service provider. For achieving good communication link among them, it is very necessary to find suitable protocol. In this paper, we discuss different hardware techniques for power monitoring, power management and remote power controlling at home and transmission side and also discuss the suitability of Zigbee for required communication link. Zigbee has major role in monitoring and direct load controlling for efficient power utilization. It covers enough area needed for communication and it works on low data rate of $20 Kbps$ to $250 Kbps$ with minimum power consumption. This paper describes the user friendly control home appliances, power on/off through the internet, PDA using Graphical User Interface (GUI) and through GSM cellular mobile phone.

 \end{abstract}

\begin{IEEEkeywords}
Zigbee, monitoring, controlling, smart grid
\end{IEEEkeywords}

\IEEEpeerreviewmaketitle
\section{Introduction}

\IEEEPARstart{i}n many countries, communication based controlling and monitoring architecture is used in smart grid to save power. Communication network may be wired or wireless. Communication through wired interface is very intricate and hard to implement or install. Wireless interfaces are chosen because they are easy to organize and install. Furthermore, Zigbee has some technical advantages over bluetooth, WiFi, infrared rays etc. Zigbee is a kind of low power-consuming communication technology for coverage area surrounded by $200 m$, with a data rate ranging from $20 Kbps$ to $250 Kbps$, it is appropriate for use in home area networks, mainly for the remote control of electric home appliances. Table. 1 shows the comparisons of bluetooth, WiFi, Zigbee.

\begin{table}[!h]
\centering
   \begin{tabular} { | p {1cm}| p {1.1cm}| p{1.2cm}| p {1.1cm}| p {1.2cm}| p{.6cm}|}
   \hline
    \textbf{Standard} & \textbf{Range} & \textbf{Number \newline of Nodes} & \textbf{Frequency \newline Band} & \textbf{Data \newline Protection }& \textbf{Power \newline use} \\ \hline
    \textbf{Bluetooth} & 10 m & 8 & 2.4GHz & 16 bit CRC & high  \\ \hline
    \textbf{Wi-fi} & 100 m & 32 & 3.1-10.6 GHz & 32 bit CRC & high \\ \hline
    \textbf{Zigbee }& 10-200 m & More than 25400 & 868/915 Mhz 2.4 GHz & 16 bit CRC & low  \\ \hline
    \end{tabular}
    \caption{Comparisons of bluetooth, wifi, zigbee.}
\end{table}

In this paper, we discuss different options of hardware technique for power controlling and monitoring architecture. For Monitoring, hardware is based on current or voltage measuring circuits, Micro Controller Unit (MCU) relay and Zigbee Reduce Function Device (RFD). Current/voltage measuring circuit measures the I/V and sends the information to MCU. Micro Controller checks abnormality of power and send the information to the home server where database is maintained through Zigbee RFD. For controlling purpose, relay is added in power monitoring hardware. In case of emergency found by MCU, relay cuts the power supply to the electric home appliances after receiving the control command. Graphic User Interface (GUI) software is used as an interface between user and end devices. User can control all electric appliances through cell phone, computer or laptop.

\begin{figure*}[t]
\centering
  \includegraphics[height=7cm,width=12cm]{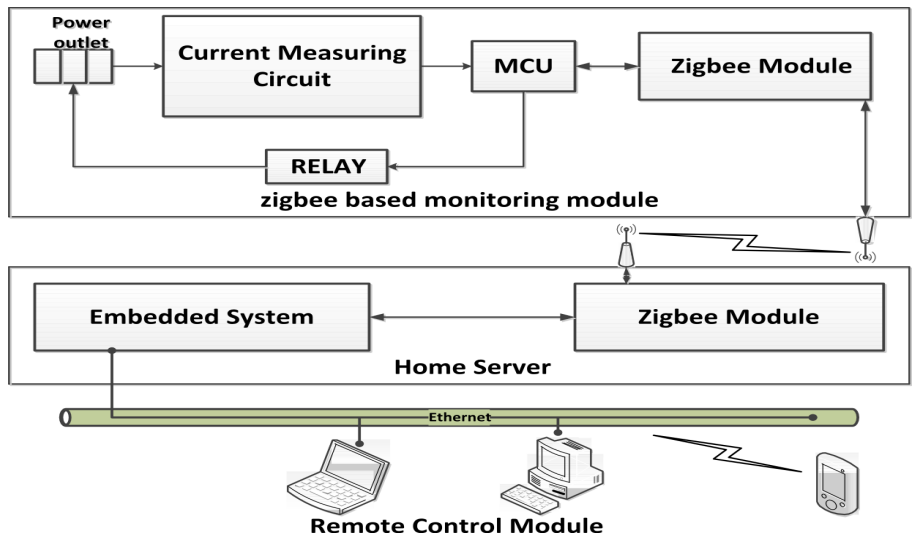}
  \caption{Power Management Structure}
\end{figure*}

\section{Related Work and Motivation}
For remote power control and monitoring, many wireless technologies are discussed; these are: Infrared rays, WLAN, Bluetooth, and Zigbee. However, Zigbee is suitable for remote power control and monitoring due to large coverage area up to $200 m$ and with transmission rate ranging from $20 Kbps$ to $250 Kbps$. Typically home appliance has three power modes, normal, stand by and off. In normal mode, home appliances use maximum power, however, in stand by mode, home appliances use almost $10$ percent of  power consumption as compared to normal mode. In [1], authors discuss architecture of automatic cut-off for home appliances during stand by mode to save power consumption.

In [2], [3] and [4], power monitoring module calculates power consumption of electric home appliances in normal mode. Zigbee communication module is used to send measured data of current and voltage to server module and store it in computer. Embedded board is used instead of computer to reduce power consumption. In case of overload/abnormality in power consumption, server module sends a control message to MCU via zigbee modul. MCU cuts power to the load  for safety purpose. Authors design the GUI by using Visual Basic (VB) to provide user friendly environment. However, power supply to power monitoring and controlling component is not discussed by authors.

In [5], Bluetooth is used for controlling electric home appliances. Bluetooth has low coverage area up to $10 m$. Bluetooth is better option for short distance controlling, however, for long distance controlling is not easy to achieve via bluetooth.

In [6], author discuss use of Zigbee communication in Advance Metering Interface (AMI). Zigbee is used to transmit detailed information of power consumption and also discuss the way of joining Zigbee network.

\section{Hardware}
In this section, we discuss different hardware options with different techniques for power monitoring and control purpose in smart grid.

\subsection{Power Management Structure}
In [3], power management structure for electric outlet as shown in Fig. 1, has three main parts: Zigbee based power monitoring module, home server and remote control section. Power monitoring module based on (MCU), Solid State Relay (SSR), a current measuring circuit and a Zigbee End Device (ZED). The home sever module has two parts: one is Zigbee Coordinator (ZC) and other is embedded board. The remote control section can be a personal Computer (PC) or Personal Digital Assistant (PDA).

\begin{figure*}[t]
\centering
  \includegraphics[height=10cm,width=12cm]{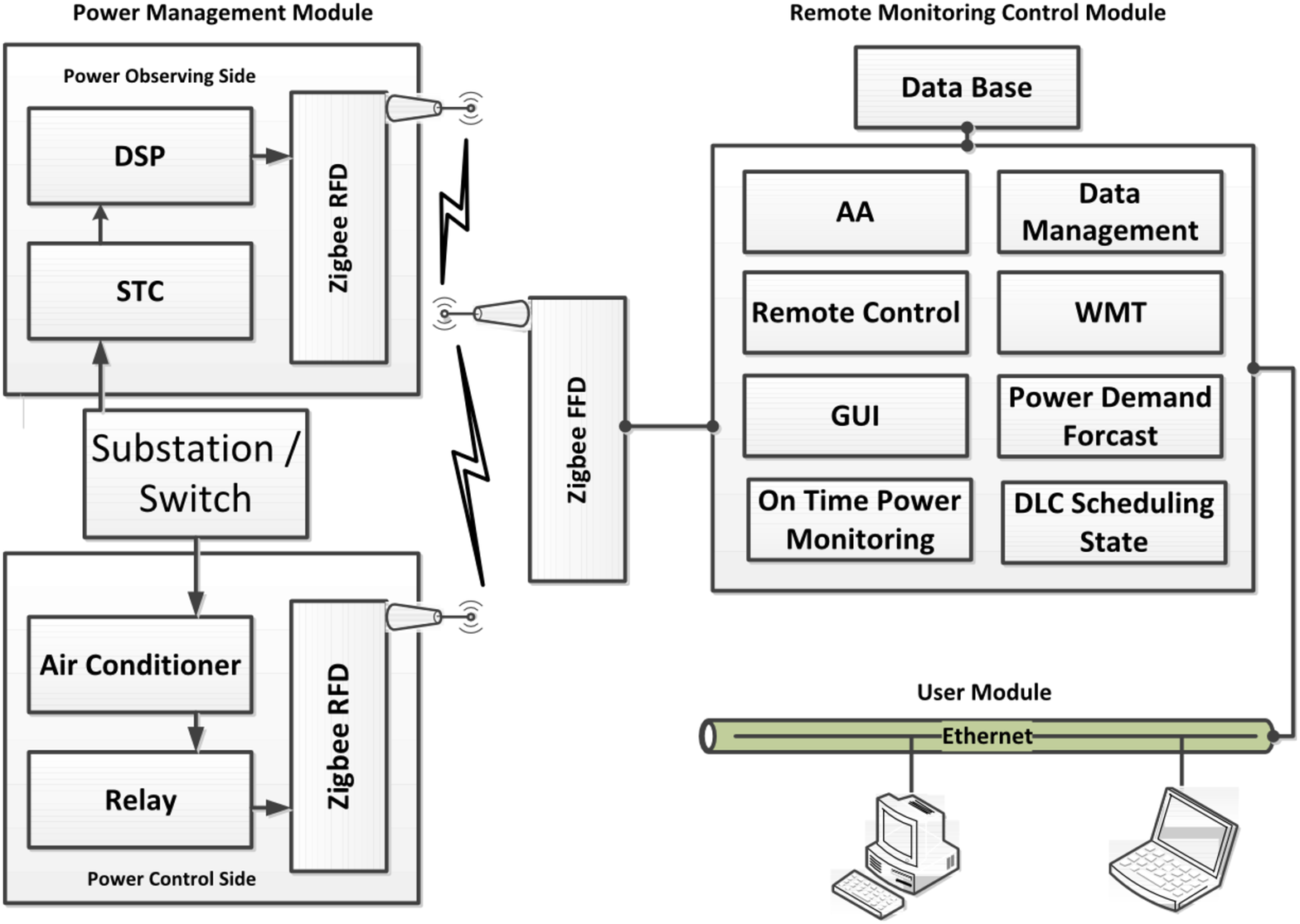}
  \caption{Real-time Power Monitoring Architecture with Direct Load Control}
\end{figure*}

Sensor in Zigbee based power monitoring module senses electric current being used by electric outlet. Current measuring circuit measures current and sends to MCU. MCU turns the electric outlet on/off. MCU based on two-way control mechanism: one direction for on/off control and other for current measurement. MCU handles ID of each end device, control electric outlets and send control message to home server through Zigbee coordinator. Through MCU, we can detect condition of all electric home appliances. Embedded system stores data of all power outlets and serve the user when they want.

To minimize power consumption, traditional relay is replaced by SSR for controlling on/off power outlets. In Zigbee based monitoring module, embedded board is used instead of PC to minimize the size, power consumption, and cost of home server.

\subsection{Real-time Power Monitoring Architecture With Direct Load Control}
In [4], Zigbee based real-time power monitoring architecture consists of three modules: power management module, Remote Monitoring Control Module (RMCM) and user module. Power management module contains Power Observing Side (POS) and Power Control Side (PCS). POS monitors power consumption from various facilities like power consumption through switchboard and substation. Alternatively, large power utilization facilities such as central air conditioner, controlled by PCS. RMCM enclose major software components of POS and database. ZC is linked to RMCM through RS-232 interface for sending and receiving messages to Zigbee device on power management module. On user side, user can remotely operate POS through any web browser on notebook or PC, as shown in Fig. 2.

POS has three main parts Signal Transformation Circuit (STC), Digital Signal Processing (DSP) board and RFD. STC collects power consumption data from switch/substation and scale the information to fit input scope of Analog to Digital Converter (A/C) convertor of DSP board. DSP board is liable to calculate various power parameters and become aware of power omission in real time. This power information is transmitted to server side through Zigbee RFD. Zigbee RFD is an end device that can collect data through different sensors or switches and send to gateway (coordinator or router).

In power control, Zigbee RFD is directly connected to relay. After finding power abnormality ZC sends command to Zigbee RFD to unload assigned power facilities.

RMCS has following major functional components. Authentication and Authorization component (AA) verifying consumer identity and used for security purpose.Warning Message Transmission (WMT) component is capable of sending warning message to user through e-mails and mobile phone short message web service after finding abnormality in power. GUI component provides different user-friendly graphical web interface to control system through general browsers. Through remote control component, user can manually perform Direct Load Control (DLC). Data management component manages different data in system like user data, historical data of power consumption. The power demand forecasts component forecast the best possible contract capacity based on historical power consumption data.

\subsection{Power Observing Structure using Bluetooth, Ethernet and GSM }
In [5], Bluetooth Controlled Power Outlet Module (BCPOM) is hardware architecture for power monitoring and controlling, as shown in Fig. 3. This hardware consists on several AC power sockets, vital control side, Protected Digital Card (PDC) side, Scalable Source Routing (SSR), power measuring side and  communication interfaces based on Zigbee and GSM. SSR is used to turn on/off each sokets where electric home machines/appliances can be pluged.

Vital control side is based on four functions: \textit{(i)} dealing out commands from communication interfaces bluetooth, ethernet and GSM, \textit{(ii)} scheming the SSR on/off, \textit{(iii)} observing the status of electric home appliances, and \textit{(iv)} transmitting the power status and measured data to the protected Digital Card Side. GSM interface is used to connect vital control side and also allows to make a call by using GSM network. Bluetooth side is a low power embedded bluetooth with a built-in high-output antenna.

The ethernet side connects electric home machines to internet. PDC side is used for storing the measurement data of electric home machines and also store status of these appliances.

\begin{figure*}[t]
\centering
  \includegraphics[height=7cm,width=10cm]{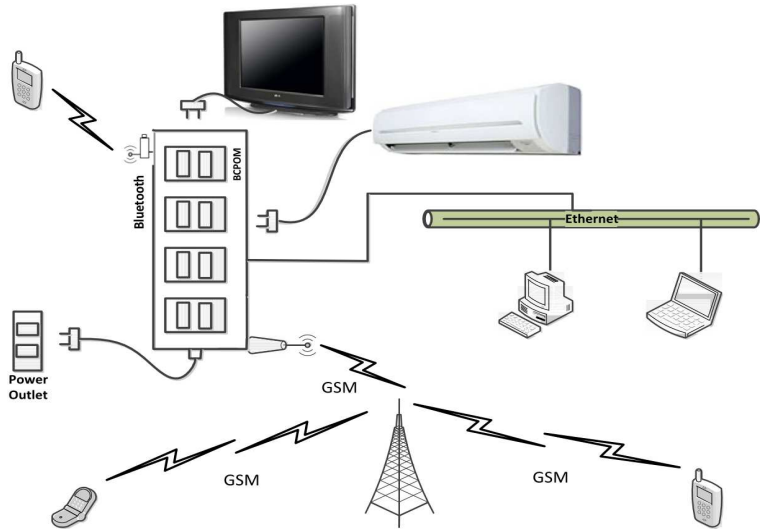}
  \caption{Power Observing Structure via Bluetooth, Ethernet and GSM}
\end{figure*}

\subsection{Power Monitoring and Controlling Architecture}
In [2], proposed system for power monitoring is based on three components: Data Acquisition (DA), Data Processing (DP) module and Application. Data acquisition consist of wireless sensors which are used for measuring AC and control power outlets. A wireless sensor has three technical parts, MCU with A/C, a sensing unit and a Zigbee transceiver. DP module collects information from all sensors through Zigbee and make a database of all collected information and responds to appeal from users.

\subsection{Power Controlling Architecture}
Authors in [1] discuss the different techniques for controlling power in different conditions. One is automatic power cut off during standby mode. Architecture of automatic power cut off during standby condition of power outlet is based on A/C conversion, a two port relay, a micro controller and power observing circuit. AC input is connected to two port relay. One port of relay is directly connected to AC output electric power outlet and other relay is connected to the output relay through power observing circuit. Power observing circuit has three major components: transformer, rectifying diodes and additional components. In micro controller unit Zigbee RF module is added to communicate with the coordinator.

 \begin{figure*}[t]
\centering
  \includegraphics[height=6cm,width=10cm]{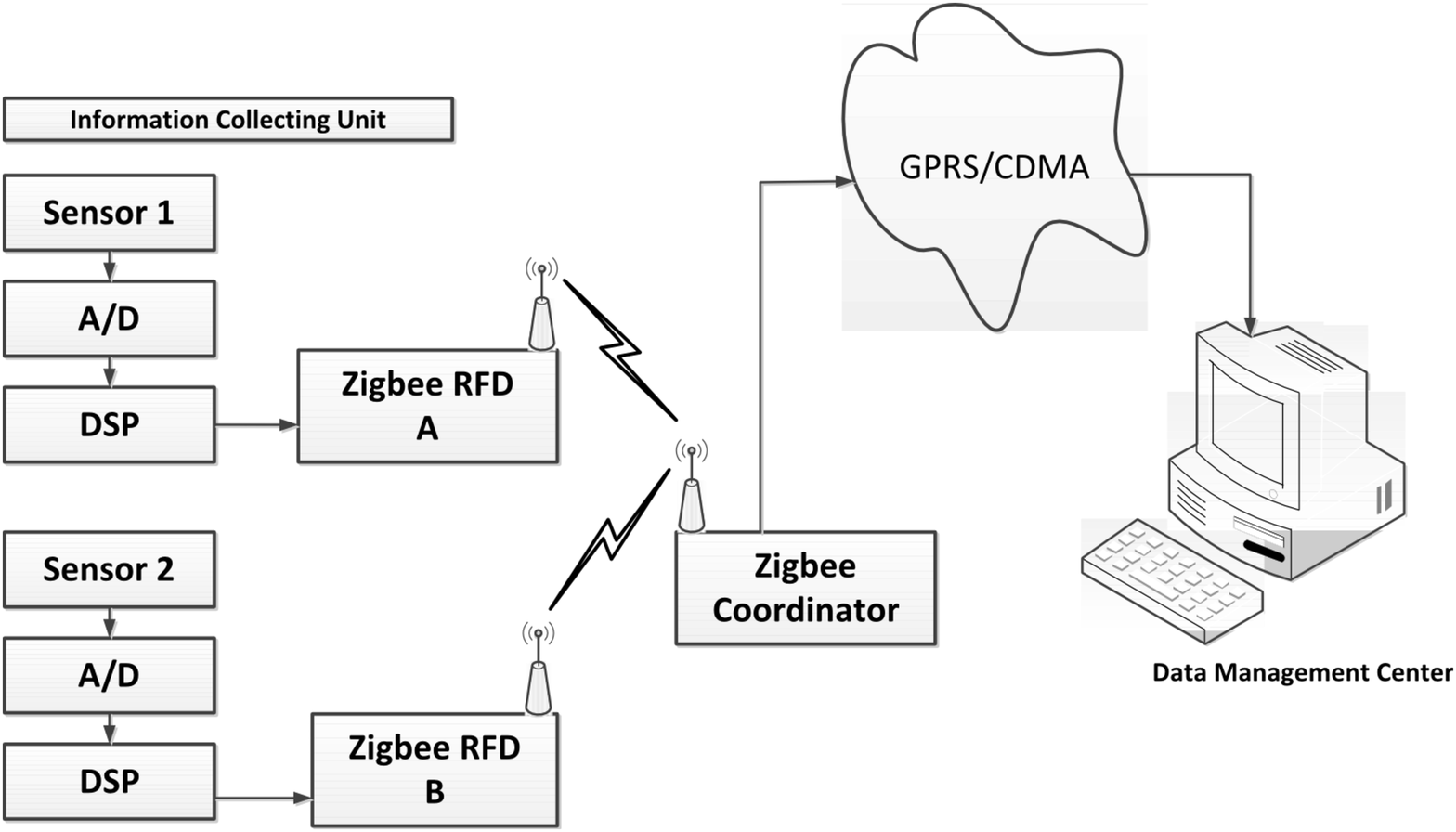}
  \caption{Zigbee-based Transmission Line Monitoring}
\end{figure*}

Other is Zigbee controller with IR code Learning Functionality. ZC is based on A/C conversion circuit, a MCU with Zigbee RF module, IR receiver and several switches. These several switches have different tasks. First switch is assigned to the first power outlet; through this button first power outlet will be controlled. Next two buttons can be assigned for the light control. One button is for `dark' function and other is for `light' function.

\subsection{Zigbee Based Transmission Line Monitoring}
In [7], authors propose a Zigbee based transmission line monitoring system. The block diagram of the system is shown in Fig. 4. This system is based on Information Collection Unit (ICU), ZC and data management center. Information collection unit consists of sensor A/C, DSP, Zigbee RFD. Sensors keep information of power consumption in analog form and A/C convert analog information into digital information and send to DSP unit. DSP unit calculates variation in power consumption or any abnormality in power utilization and send the information to ZC through Zigbee RFD. ZC sends all information coming from different sensor to data management center.

\section{Suitability of IEEE802.15.4 for Smart Grid}
There are various  types of wired and wireless interfaces that can be used in smart grid for communication purpose similar to Bluetooth, WiFi, LAN, Zigbee, Ethernet, IEEE 1394, PLC, etc. Communication through wired interface in smart grid is very intricate and hard to implement and install; that is why wireless interfaces are chosen, since they are easy to organize and install. Furthermore, Zigbee has some technical advantages over others. It is low-cost, low power consuming wireless communication standard with maximum number of nodes. Zigbee has quicker response time. In [6], a new Zigbee based  power meters can easily join the Zigbee network. Power meter sends the date and time synchronization command to the Zigbee device which sends this command to ZC. ZC replies with date and time data to the Zigbee device after completing the date time synchronization power meter ready to send the required data and become a part of Zigbee network. Below we tell the Open System Interconnection (OSI) model layer on which Zigbee protocol based and Zigbee based power monitoring and controlling.

\subsection{Zigbee}
In [8], mound of Zigbee protocol is based on four layers of standard Open System Interconnection (OSI) model. Application Layer, Network Layer, Medium Access Control Layer (MAC) and Physical Layer. Application Sub layer layer is in charge for maintenance of required tables that match two devices according to their desires and services. Network layer is responsible for creating network by adding and deleting the nodes and for the route discovery between devices and their maintenance. Zigbee has two standards for physical layers that work in different frequency range: 868 MHz, 915 MHz and 2.4 GHz. MAC layer is conscientious to offer interface between the Service Specific Convergence Sub layer (SSCS) and physical layer.

\subsection{Zigbee Based Monitoring and Controlling}

 \begin{figure*}[t]
\centering
  \includegraphics[height=8cm,width=8cm]{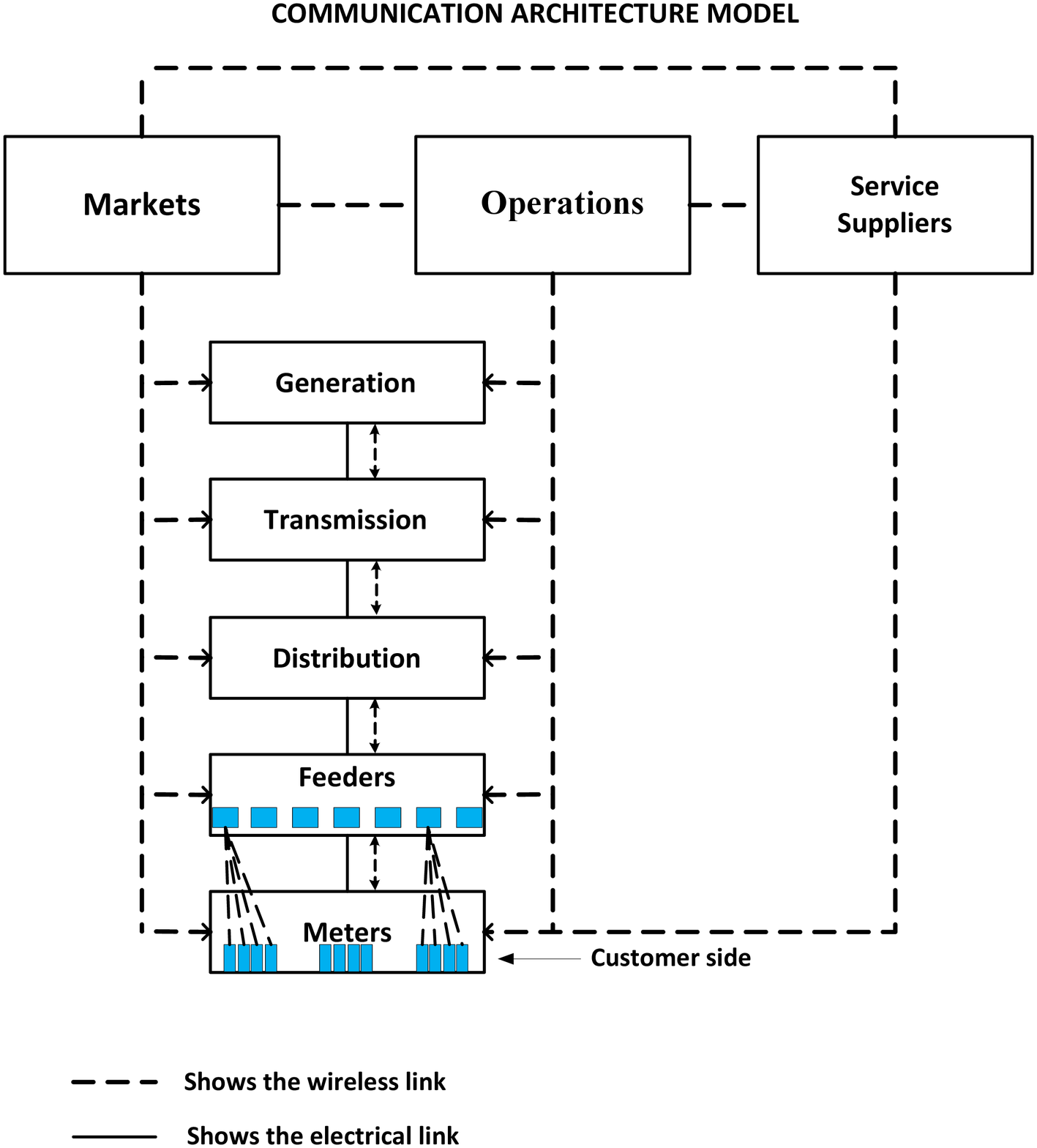}
  \caption{Communication Architecture}
\end{figure*}

Zigbee technology can be implemented in generation, transmission, distribution, and in consumer sectors to provide control on power consumption and give accurate data to user and utility. Fig. 5 shows communication architecture, where generation side, transmission side, distribution side and consumers are not only connected electrically but also connected by communication interfaces with markets, operations and service providers. Zigbee based devices have two types: one is Full Function Device (FFD) and other is RFD. FFD can be coordinator or router. It is liable to create the network, select the radio frequency channel and distinctive network identifier to avoid collision among data from different sensor nodes and for security purpose. Zigbee RFD plays role of end device that collects variety of data from different switches and sensors. Here we discuss Zigbee based transmission lines monitoring, meter reading system and load control system.

\subsubsection{Zigbee Based Transmission Lines Monitoring}

In [7], transmission line monitoring system is based on environmental parameters of lines and towers such as temperature, ice, wind lightening etc. Sensors are placed on transmission line. These Zigbee based sensors sense power passing through it and send to data center via Zigbee gateway that are fixed on transmission tower. Small blue boxes represent power monitoring device as shown in Fig. 6.

\begin{figure*}[t]
\centering
  \includegraphics[height=8cm,width=10cm]{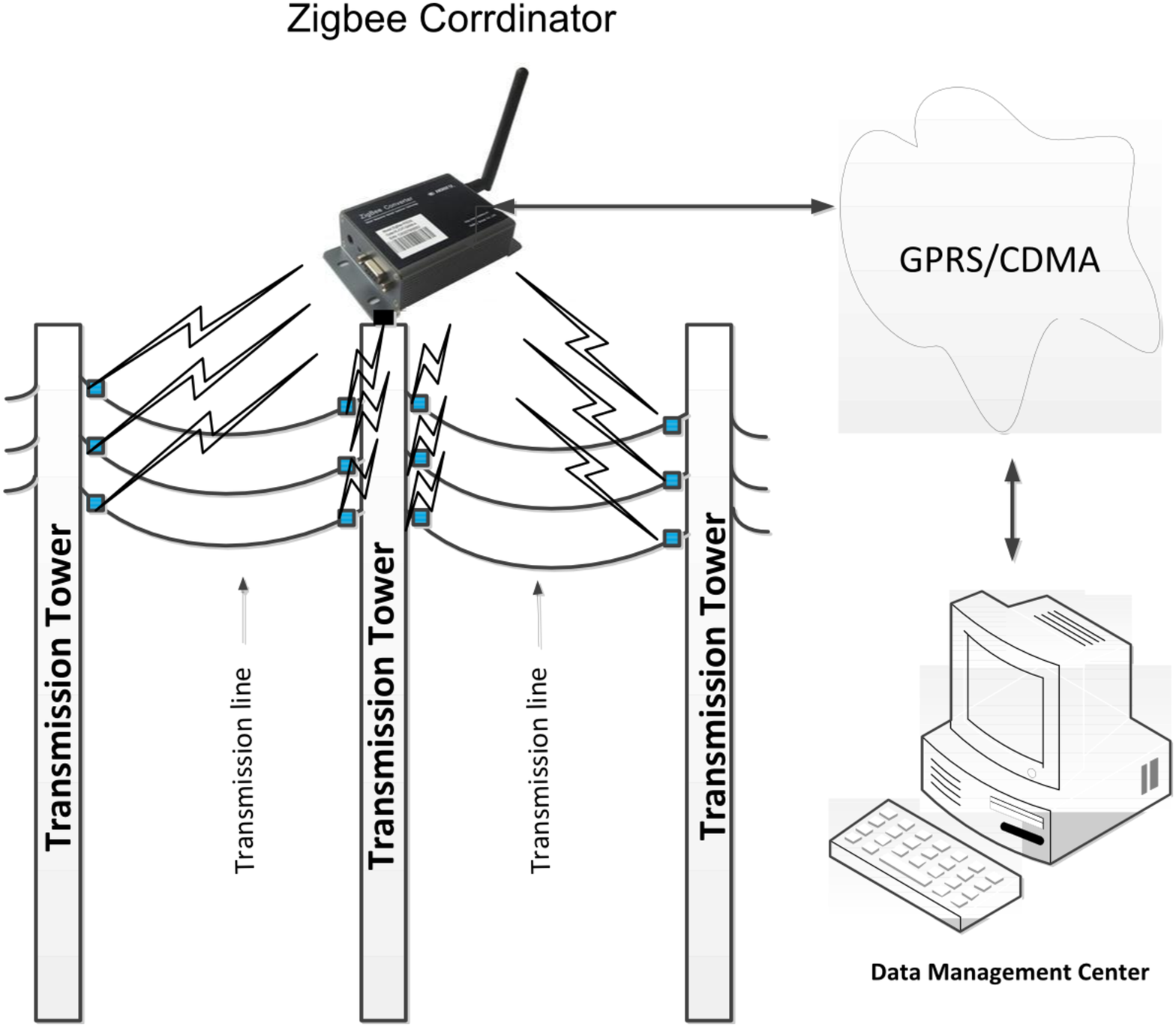}\\
  \caption{Transmission Line Monitoring}
\end{figure*}

\subsubsection{Zigbee Based Meter Reading System}

In [8], Zigbee based meter reading system has two main parts: Zigbee network and database unit. Zigbee network is constructed by Zigbee router, end devices and coordinator. Router and coordinator must be FFDs and end devices can be RFDs. After calculating required information end devices send this information to the relevant router. Router has function to synchronize service to other FFDs like coordinator. Coordinators send received data through $RS232$ interface to database unit.

\subsection{Zigbee Based Load Control System}

In [1,4], for controlling purpose, Zigbee RFD is directly connected to relay. Relay is used for turning the switch on and off. After finding any abnormality in power used by electric home appliances control command is sent to the Zigbee RFD by coordinator. Subsequently, relay turns off those home appliances. ZC also receives the data from the end devices through the zigbee RFD.

\section{User responsive control system}
Zigbee based Smart meters are part of AMI which give all information about power to customer like showing peak hours time, current running price, and real time power consumption by electric home appliances. It involves user to control power load at his end. This system provides reliable control system to user through different ways.

\subsection{Control by Internet}
In [5], the software for remote controlling is install in PC or laptop and connect to internet. The network formation of connecting remote control by internet is a host-client structure. This remote control software sets IP address and ports and sends request to Power Monitoring System (PMS), creates a link between user and PMS. In [4], when PMS detects a power abnormality, warning message hurl to user through e-mail by means of Mail Transfer Protocol (MTP). In [2], under the proposed system, functional schedules are implemented on java Virtual Machine(VM). When user want to switch on/off an electric home appliance from a web interface, response time is less than one sec.

\subsection{Control by PDA with GUI}
In [1], [3], [4] and [5], PDA is connected to power monitoring module for controlling and monitoring electric home machines. GUI is used in control area to create interface between user and electronic devices. By making use of GUI, user can access condition of electric home appliances, time reaction and energy utilization made by electric power outlets. GUI provides effortless control of power status of electric home machines. User can set each switch as on/off by sending command. The remote control system offers to supervise and manage power condition of electric home appliances.

\subsection{Control by GSM Cellular Mobile Phone}
In [5], consumer can examine and organize electric home machines using GSM cellular mobile phone. User sends control command message through GSM cellular mobile phone to PMS. After receiving control command PMS allows user to control if sending command format match with system. User can be capable for supervise and organize power condition of electric home appliances anytime and anywhere using GSM cellular mobile. PMS can also send warning message to consumer after finding unusual status of power

\section{Conclusion}
In this paper, we discussed power utilization, power organizing and power controlling architecture for power saving purpose. We also discussed the role of Zigbee in transmission line monitoring, real time meter reading and direct load controlling of electric home appliances. This paper also describes the user friendly control home appliances for power on/off through internet, PDA using GUI and through GSM cellular mobile phone.

\end{document}